\documentclass[12pt, a4paper, fleqn]{article}
\usepackage{amssymb}
\usepackage{amsmath}
\usepackage{graphics}
\usepackage{graphicx}
\usepackage{bm}
\usepackage{geometry}
\usepackage{dsfont}
\usepackage{arydshln}
\usepackage{xcolor}
\usepackage{setspace}
\usepackage{hyperref}
\usepackage{natbib}
\usepackage{rotating}
\usepackage{authblk}

\numberwithin{equation}{section}

\title{\textbf{Familywise error rate control for block response-adaptive randomization}}

\author[a,b]{Ekkehard Glimm}
\author[c]{David S.\ Robertson}

\affil[a]{Novartis Pharma AG, Basel, Switzerland}
\affil[b]{University of Magdeburg, Medical Faculty, Institute of Biometry and Medical Informatics, Germany}
\affil[c]{MRC Biostatistics Unit, University of Cambridge, UK}

\date{\vspace{-30pt}}

\begin{document}

\maketitle

\doublespacing

\begin{abstract}
Response-adaptive randomization allows the probabilities of allocating patients to treatments in a clinical trial to change based on the previously observed response data, in order to achieve different experimental goals. One concern over the use of such designs in practice, particularly from a regulatory viewpoint, is controlling the type~I error rate. To address this, Robertson and Wason (Biometrics, 2019) proposed methodology that guarantees familywise error rate control for a large class of response-adaptive designs. In this paper, we propose an improvement of their proposal that is conceptually simpler, in the specific context of block-randomised trials with a fixed allocation to the control arm. We show the modified method guarantees that there will never be negative weights for blocks of data, and can also provide a substantial power advantage in practice.\\[-12pt]

\noindent \textbf{Keywords:} conditional invariance principle, multiple testing, type~I error rate
\end{abstract}

\section{Introduction}

Randomized clinical trials are typically designed in such a way that a decision about treatment efficacy is reached as quickly as possible and with a minimum number of patients exposed to inferior treatment options. Response-adaptive randomization (RAR) can help achieve such goals by an allocation process that makes randomization of a newly recruited patient dependent on responses to treatment from previous study participants. This can offer advantages in terms of the benefit for patients recruited into the study, or the study's power to detect treatment effects. Many different classes of RAR procedures have been proposed for various trial contexts, and a recent review of methodological and practical issues around the use of RAR in clinical trials can be found in Robertson et al.~\cite{robertson2020response}. However, one major concern is about the type~I error rate arising from such studies. This is particularly the case from a regulatory viewpoint, where control of the type~I error rate is required for confirmatory studies \citep{FDA2019, European2002}.

\cite{Robertson2019} proposed methodology that guarantees type~I error control by iterative application of the conditional invariance (CIV) principle. For that purpose, they assume the existence of an auxiliary design in which the test statistic has a known null distribution. At interim analyses of the RAR trial (this may be after blocks of patients, or after every patient), the randomization ratios may be changed. The test statistic used to test for treatment efficacy, however, will be calculated in such a way that its null distribution matches the null distribution of the test statistic from the auxiliary design. \cite{Robertson2019} show that this guarantees type~I error control at level $\alpha$ if the test in the auxiliary design is a level $\alpha$ test. For multi-arm trials, the testing procedure controls the familywise error rate (FWER), which is defined as the maximum probability of at least one type~I error under any configuration of true and false null hypotheses.

In this paper, we present an improvement of their proposal based on the CIV principle. It is simpler in that it requires only matching the variance of the quantities that form the final test statistic, rather than matching both variance and means. This restricts the method to block-randomization and to a fixed randomization to the control arm. On the other hand, by construction the method guarantees that there will never be negative weights for blocks of data and can also provide a substantial power advantage.

The outline of the rest of the paper is as follows. In Section~\ref{sec_matching} we describe the proposed testing procedure and its connection with existing approaches. A simulation study is presented in Section~\ref{sec:sims} to compare the testing strategies, and we conclude with a discussion in Section~\ref{sec:discuss}.

\section{Matching algorithm}\label{sec_matching}

Consider a trial with $K > 1$ experimental treatment arms and a control arm. We assume that the control arm is not subject to any response-adaptive randomization. For the treatment arms, there is a run-in period which allocates patients according to a predefined randomization scheme. After the end of this run-in, randomization to the experimental treatment arms may be modified based on previous results. The aim is to `reweight' contributions to the test statistic in such a way that the FWER is controlled.

Let $i=1,\ldots,n_0^*$ be the patients randomized in the run-in period. Subsequently, RAR is used for patients $i=n_0^*+1, \ldots, n^*$.
For simplicity, we consider only the test of the null hypothesis $H_1: \mu_1=\mu_0$, where $\mu_1$ and $\mu_0$ are the expected responses from patients under treatment and control, respectively. Extensions to intersection hypotheses within a closed test procedure will be discussed in Section~\ref{subsec:CTP}.
We assume that $X_i\sim N(\mu_1,\sigma^2)$ is the response obtained from patient $i$ randomized into treatment group~1. Without loss of generality, we assume $\sigma^2=1$. To keep notation simple, we will not introduce notation and labelling of patients in the other treatment groups here. We assume that $n_0$ patients are randomized to treatment 1 in the run-in period.

Furthermore, we assume the existence of an auxiliary design, which can be thought of as one of the allowed randomization lists, chosen before the beginning of the trial. Unless there are reasons to choose otherwise, a default option would be to use fixed (equal) randomization to reflect the uncertainty before the trial begins over which of the treatment options will be superior. Like in \cite{Robertson2019}, $Y_i\sim N(\mu_1,1)$ denotes the random variable for the `intended' response of a patient in treatment group~1 according to the auxiliary design that foresees a total of $n$ patients in treatment group~1.
For $i\leq n_0$, we have $Y_i=X_i$. Afterwards, the two designs diverge.

In this setup, \cite{Robertson2019} describe a method which uses the resulting data to calculate a test statistic $\tilde{T}$ for the test of $H_1$. The test statistic is a difference of weighted averages of the observations on treatment arm~1 and the control arm. The weights are calculated recursively based on the number of allocations to the experimental treatments and the control.

We now describe a simplification of the handling of the control group in this setup. We ignore the control group in the updating steps of the algorithm described in the Web Appendices~B and~D of \cite{Robertson2019}. This implies that no interim tests for efficacy are performed in this simplification. As the subsequent description will show, this assumption can be relaxed for the blockwise RAR if we fix the number of control patients obtained per block.
The basic idea of the modified proposal is to use the `matching algorithm' only for the variance, thereby allowing shifts in the mean. Due to this, the CIV principle may no longer be used as a justification for the algorithm, but a minor modification of it still applies.

Let $\bar{y}_1(n_0)\sim N(\mu_1,\frac{1}{n_0})$ be the mean response in treatment group 1 after the run-in period. For the next block of recruited patients, the allocation rule for treatment 1 versus all other experimental treatments is changed in some way such that at the end of this block (block 1, say), we have $\tilde{n}_1$ patients in treatment arm 1 instead of the pre-planned $n_1$ from the auxiliary design. We proceed in this fashion up to a final block $b$. In every block, we assume that at least one patient will be randomized to treatment~1. The following formulae define some resulting summary statistics and their conditional distributions given the observed sample sizes:

\begin{itemize}
\item Block 0: This is the run-in period. After this block, we obtain
$
T_0=\frac{n_0}{n}\bar{y}_1(n_0)\sim N\left(\frac{n_0}{n}\mu_1,\frac{n_0}{n^2}\right)
$
and set $w_0=n$.

\item Block 1: We define \vspace{-12pt}
\begin{eqnarray*}
&&T_1=\frac{n_1}{w_0}\bar{y}_1(n_1)\sim N\left(\frac{n_1}{w_0}\mu_1,\frac{n_1}{w_0^2}\right)\\
&&\tilde{T}_1=\frac{\tilde{n}_1}{w_1}\bar{x}_1(\tilde{n}_1)\sim N\left(\frac{\tilde{n}_1}{w_1}\mu_1,\frac{\tilde{n}_1}{w_1^2}\right)\\
&&T_{(1)}=\frac{n_{(1)}}{w_0}\bar{y}_1(n_{(1)})\sim N\left(\frac{n_{(1)}}{w_0}\mu_1,\frac{n_{(1)}}{w_0^2}\right)\\
&&\tilde{T}_{(1)}=\tilde{T}_1+T_{(2)}
\end{eqnarray*}
Here, $n_{(k)}=n_k+\ldots+n_b$ denotes the sample sizes of treatment group~1 in the blocks of the auxiliary design and $T_{(k)}$ the corresponding weighted sum of expected values; $\bar{y}_1(n_1)$ is the average response of block~1, treatment~1 in the auxiliary design; $\bar{x}_1(\tilde{n}_1)$ the average response of actually observed observations in block 1, treatment 1. Since $\tilde{n}_1$ is a function of the entire data from patients $1,\ldots, n_0^*$, the given distributions of $\tilde{T}_1$ and $\tilde{T}_{(1)}$ are conditional on the block~0 data.
We now calculate $w_1$ in such a way that the variance of $T_{(1)}$ and that of $\tilde{T}_{(1)}$ are matched. Hence,
$
w_1=n\cdot\sqrt{\frac{\tilde{n}_1+n_{(2)}}{n_1+n_{(2)}}} \; ,
$
leading to the conditional distribution given the run-in-phase-data as
$
\tilde{T}_{(1)}\sim N\left(\frac{\tilde{n}_1+n_{(2)}}{w_1}\mu_1,\frac{n_{(1)}}{n^2}\right).
$
\item Block $k$:
After every block, randomization maybe changed. For block $k$ we define: \vspace{-12pt}
\begin{eqnarray*}
&&T_k=\frac{n_k}{w_{k-1}}\bar{y}_1(n_k)\sim N\left(\frac{n_k}{w_{k-1}}\mu_1,\frac{n_k}{w_{k-1}^2}\right)\\
&&\tilde{T}_k=\frac{\tilde{n}_k}{w_k}\bar{x}_k(\tilde{n}_k)\sim N\left(\frac{\tilde{n}_k}{w_k}\mu_1,\frac{\tilde{n}_k}{w_k^2}\right)\\
&&T_{(k)}=\frac{n_{(k)}}{w_{k-1}}\bar{y}_1(n_{(k)})\sim N\left(\frac{n_{(k)}}{w_{k-1}}\mu_1,\frac{n_{(k)}}{w_{k-1}^2}\right)\\
&&\tilde{T}_{(k)}=\tilde{T}_{k}+T_{(k+1)}
\end{eqnarray*}
with
$
w_k=w_{k-1}\cdot \sqrt{\frac{\tilde{n}_k+n_{(k+1)}}{n_k+n_{(k+1)}}} \, .
$
Consequentially, the conditional distribution given run-in-phase and blocks $1, \ldots, k-1$ is:
$
\tilde{T}_{(k)}\sim N\left(\frac{\tilde{n}_k+n_{(k+1)}}{w_k}\mu_1,\frac{n_{(k)}}{n^2}\right).
$

\item Final block $b$: We define \vspace{-12pt}
\begin{eqnarray*}
&&T_b=\frac{n_b}{w_{b-1}}\bar{y}_1(n_b)\sim N\left(\frac{n_b}{w_{b-1}}\mu_{b-1},\frac{n_b}{w_{b-1}^2}\right)\\
&&\tilde{T}_b=\frac{\tilde{n}_b}{w_b}\bar{x}_b(\tilde{n}_b)\sim N\left(\frac{\tilde{n}_b}{w_b}\mu_1,\frac{\tilde{n}_b}{w_b^2}\right)
\end{eqnarray*}
Since no additional block follows, we have
$
w_b=w_{b-1}\sqrt{\frac{\tilde{n}_b}{n_b}},
$
such that
$
\tilde{T}_b\sim N\left(\frac{\tilde{n}_b}{w_b}\mu_1,\frac{n_b}{n^2}\right).
$
\end{itemize}

%{\color{red} We could shorten this description to the description of only step $k$ by defining $T_{(b+1)}=0$ and $n_{(b+1)}=0$. For $w_0$, we have $w_0=n\cdot \sqrt{\frac{n_0+n_{(1)}}{n_0+n_{(1)}}}=n$.}

\noindent At the end of this procedure, we obtain the statistic
$
\tilde{T}=T_0+\sum_{k=1}^b \tilde{T}_k.
$
The corresponding statistic from the auxiliary design is $T=\sum_{k=0}^b T_k$.
By construction, both $T$ and $\tilde{T}$ have variance $\text{var}(T)=\frac{1}{n}$. However, while
$E(T)=\mu_1$, $E(\tilde{T})$ cannot be easily derived due to the response-adaptive modifications.

Since we did not modify the randomization to the control treatment $0$, we have a consistent estimate of $\mu_0$ from $\bar{z}(m)\sim N(\mu_0,\frac{1}{m})$ where $m$ is the planned number of control patients in the auxiliary design and $\bar{z}(m)$ is the final average response in the control group across all blocks. If we had known $\mu_0$ at the time of doing the blockwise-RAR, we could simply have subtracted $\mu_0$ from every observation $x_i$ and run the algorithm in the described way on $x_i-\mu_0$ instead of $x_i$. Since $E(x_i)=\mu_1=\mu_0$ under $H_1$, this would have led to $E(\tilde{T}_k)=0$ and remove the need to match the weights on the means as well. As we have a record of all weights $w_1,\ldots, w_b$, we can do this `post-hoc':
\begin{equation*}
\tilde{T}^*=T_0-\frac{n_0}{n}\bar{z}(m)+\sum_{k=1}^b \left[\tilde{T}_k-\frac{\tilde{n}_k}{w_k}\bar{z}(m)\right]=
\tilde{T}-\sum_{k=0}^b u_k \cdot \bar{z}(m),
\end{equation*}
where $u_k=\frac{\tilde{n}_k}{w_k}$ for $k=0,\ldots,b$, $\tilde{n}_0=n_0$ and $w_0=n$.
The final test statistic is then
$
\tilde{U}=\frac{\tilde{T}^*}{std(\tilde{T}^*)},
$
where $std(\tilde{T}^*)=\sqrt{\frac{1}{n}+(\sum_{k=0}^b u_k)^2 \frac{1}{m}}$.
Asymptotically, $\tilde{U}$ is distributed as $N(0,1)$ under $H_1: \mu_1=\mu_0$ since $\bar{z}(m)\sim N(\mu_0,\frac{1}{m})$.

%A way of dealing with the "finish up" problem could be to assign a last block of 3 patients, 1 for each treatment, which will not be changed anymore by the RAR.

Hence, $\tilde{U}$ is independent of any modifications to the randomizations that modified the originally planned $n_k$ within the blocks. The asymptotic condition is `mild' in the sense that for a reasonably long-running study, $\sqrt{m}(\bar{z}_m-\mu_0)\sim N(0,1)$ will hold by standard asymptotic theory for estimates of expected values (if either there are many blocks or the blocks are large), such that the concern about the normal distribution assumption is similar to concerns about estimating $\sigma^2$ from the data.

The approach can be modified to allow early stopping for efficacy if the number of control patients is fixed per block. In that case, we can setup the approach similar to a group-sequential trial with $\alpha$-spending.
Rather than calculating $\tilde{U}$ only once at the end of the trial, we would calculate $\tilde{U}_F$ at the end of a randomly selected block $F<b$ by treating $F$ as the last block and using $\bar{z}(m_F)$ instead of $\bar{z}(m)$. Since the number $m_1,\ldots,m_b$ of control observations after block $1,\ldots,b$ are fixed in advance, $\bar{z}(m_F)\sim N(\mu_0,\frac{1}{m_F})$ such that $\tilde{U}_F\sim N(0,1)$ still holds. If there is a sequence $\alpha_1,\ldots, \alpha_b$ such that $\sum_{k=1}^b \alpha_i=\alpha$, then rejecting $H_1$ when $\tilde{U}_F\geq \Phi^{-1}(1-\alpha_F)$ controls the type~I error rate by the Bonferroni inequality. Unfortunately, a `true' group-sequential approach is harder to implement since the correlation between $\tilde{U}_F$ and $\tilde{U}$ depends on $u_{F+1}, \ldots, u_b$ which are not available at the time of the interim analysis. The correlation {\it could} be calculated if after the interim analysis, the auxiliary design would be strictly followed for blocks $F+1,\ldots, b$. However, this would defeat the purpose of RAR.

A test of $\tilde{U}_F$ at level $\alpha$ at a randomly selected time $F$ would only control the type~I error rate if the selection of $F$ were stochastically independent of $\tilde{U}_1, \ldots, \tilde{U}_{F-1}$ under $H_{1}$. Obviously, calculating $\tilde{U}_k$ after every block and then deciding whether to test now or later based on the observed value would lead to a selection bias.

%With this strategy, early stopping with rejection of $H_1$ would not be feasible. I think, however, that it is easy to relax this if the number of control patients per block is also pre-fixed in the pre-plan. In that case, the average control response after any block would also have a fixed non-adapted number of placebo patients.

\subsection{Connection with existing approaches}

There is a close connection with the proposal of~\cite{Robertson2019} (and hence our modified proposal) and more traditional adaptive designs. Assume that we want to test the one-sample hypothesis $H_0:\mu=\mu_0$ adaptively. Before study start,  we plan a first interim analysis after $n_0$ patients and a weight $w_0\leq 1$ for this first step. At the interim, we calculate the test statistic $t_0=\sqrt{n_0}(\bar{y}_0-\mu_0)$ which is $N(0,1)$ under $H_0$. Subsequently, we pick a new sample size $\tilde{n}_1$ for the next stage (up to the next interim) and a weight $w_1$ for this stage. The weight and $\tilde{n}_1$ may both depend on the data from stage~0, but the restriction $w_0^2+w_1^2\leq 1$ must be obeyed. At the end of the stage, we calculate $t_1=\sqrt{\tilde{n}_1}(\bar{x}_1-\mu_0)\sim N(0,1)$ given the stage~0 data and $t_{(1)}=w_0 t_0+\sqrt{(1-w_0)^2} t_1 \sim  N(0,1)$. The next stage then combines $t_2$ and $t_{(1)}$ using $w_1$. We continue in this fashion until at some point we decide to call the final analysis. Then, at the penultimate analysis (the last interim before this final one), we spend the rest of the weight such that $\sum_{k=1}^b w_k^2=1$. Hence, the last weight $w_b=\sqrt{1-\sum_{k=1}^{b-1} w_k^2}$ is not selected anymore. The weights $w_k$ are allowed to depend on all data up to interim analysis $k-1$, so they can be iteratively reused including the previous weights.
This approach was described by~\cite{brannath2002recursive}. The proposals described in this paper and in~\cite{Robertson2019} can be viewed as an application of this general procedure, with a special way of calculating the weights and a `time horizon' (in terms of total recruited patients $n$, say) which is given from the start.

\subsection{Application within a closed test procedure}\label{subsec:CTP}

The approach described above generalizes to the application within a closed test procedure (CTP). In the CTP (\cite{marcus1976closed}), $H_j$, $j=1,\ldots, K$ is rejected if and only if all $H_J$, $J\subseteq \{1,\ldots,K\}$ with $j\in J$ are rejected at level $\alpha$ where $H_J=\cap_{j\in J} H_j$.
In order to test $H_J: \mu_j=\mu_0\; \forall j\in J$, several approaches can be considered. For example:
\begin{itemize}
    \item All observation of the experimental treatment arms in $J$ are pooled and treated as a single treatment arm to test against control treatment $0$ with the approach described in Section~\ref{sec_matching}. This is called ``closed $z$-test" in the simulations below.
    \item Assume that the approach from Section \ref{sec_matching} is used on all experimental treatment arms separately, leading to test statistics $\tilde{U}_j$, $j=1,\ldots, K$. Using $\max(\tilde{U}_j)_{j\in J}$ as the test statistic for $H_J$, $H_J$ is rejected if $\max(\tilde{U}_j)_{j\in J}\geq \Phi^{-1}(1-\frac{\alpha}{|J|})$. This is the Bonferroni-Holm method in the simulations of Section \ref{sec:sims}.
\end{itemize}
In contrast, a ``Dunnett-like" closed test (see \cite{magirr2012generalized}) is not straightforward. The marginal null distribution $\tilde{U}_j$ is $N(0,1)$, but the conditional correlation of $\tilde{U}_{j_1}$ and $\tilde{U}_{j_2}$ is
% $$
% \frac{\sum_{k=0}^b u_{k_{j_1}}\sum_{k=0}^b u_{k_{j_2}}\frac{1}{m}}{\sqrt{\frac{1}{n_{j_1}}+\left(\sum_{k=0}^b u_{k_{j_1}}\right)^2\frac{1}{m}}\sqrt{\frac{1}{n_{j_2}}+\left(\sum_{k=0}^b u_{k_{j_2}}\right)^2\frac{1}{m}}}
% $$
% which is
not independent of the sample size modifications by the RAR.

\section{Simulation studies}\label{sec:sims}

To investigate the operating characteristics of the suggested design, we use the set-up of a trial with $J = 3$ blocks (not including the run-in), with block sizes (40, 40, 40) for all of the experimental treatments and (20, 20, 20) for the control. In the run-in period, five patients are allocated to each of the treatments including the control. We set the true control mean $\mu_0 = 0$, and~$\alpha = 0.05$. The auxiliary designs in all scenarios were simply random draws from a discrete uniform distribution on $\{1, \ldots, K\}$ where $K$ is the number of experimental treatment arms.

\subsection{Bayesian Adaptive Randomization}\label{sec_BAR}

We compare the methods under a Bayesian Adaptive Randomization (BAR) scheme.
Following~\cite{Robertson2019}, we use a similar block-randomized BAR scheme to the one in~\cite{Wason2014}. The randomization probabilities $(\pi_1, \ldots, \pi_K)$ for the experimental treatments at the~$(j+1)$th stage are given by $
\pi_i = \frac{P(\mu_i > \mu_0 \mid \bm{X} = \bm{x})^{\gamma}}{\sum_{k=1}^K P(\mu_l > \mu_0 \mid  \bm{X} = \bm{x})^{\gamma}}$,
where $P(\mu_i > \mu_0 \mid \bm{X})$ is the posterior probability that $\mu_i$ is greater than $\mu_0$ given the observed data~$\bm{x}$, see the Supporting Information for full details. In our simulations, we set $\gamma = 0.5$. As well, since our proposal requires there to be at least one patient allocated to each experimental treatment per block, we ensure this is the case by allocating the last~$K^* \geq 0$ patients in each block to each of the~$K^*$ experimental treatments that have zero observations.

\subsection{Error Inflator scheme}

To assess the FWER and power in a situation where type~I control is known to be violated, we also investigate the allocation scheme presented in Section~2.3 of \cite{Robertson2019}, adapted to block randomization. This rule keeps on allocating patients to treatment~1 (apart from one patient per block to each of the other experimental treatments) as long as the mean response of treatment~1 remains below a fixed threshold of~0.5. As soon as the fixed threshold is crossed, all subsequent patients not randomized to control are allocated with equal probability to the other experimental treatments (except for one patient per block on treatment $1$). Full details are given in the Supporting Information.

\subsection{Examples of weights}

Tables \ref{tab:example1} and \ref{tab:example2} show the weights from two simulations under BAR and the error inflator scheme, respectively. Throughout, we set $\mu_1 = 0, \mu_2 = 1$ for the experimental treatments. The proposal from \cite{Robertson2019}, denoted RW, is compared with that from Section \ref{sec_matching}.

In the BAR example, we see that there is hardly any difference between the weights produced by the two methods. They are also very similar to the actually observed sample sizes, as would be expected from the simulation setup with two equally efficacious treatments and a ``neutral" prior.

In the error inflator example, the weights are very different for the two methods and also differ from both the observed sample sizes and the sample sizes in the auxiliary design. Both weight calculations up-weight the few responses from treatment~1 and down-weight the many from treatment~2. The weight calculation from Robertson and Wason produces negative weights for the control group here -- something that is not possible with the calculation from Section~\ref{sec_matching}.

Note that the statements refer to a single simulation run. For the error inflator scheme, this is a case where treatment~1 crossed the threshold after block~0. Observed sample sizes and weights are very different for other simulation runs where this does not happen. In contrast, for the BAR scheme, simulation runs are more similar to each other in stochastic tendency. \\

% \begin{table}[ht!]
%     \centering
%     \begin{tabular}{l l l l l}
%         \textbf{Experimental treatment~1} \\ \hline
%         RW test statistic & 4.23\\
%         New test statistic & 4.20 \\ \hdashline
%         RW experimental treatment weights &  62.19 & 63.47 & 75.21 \\
%         New experimental treatment weights & 62.45 & 63.32 & 70.79 \\ \hdashline
%         RW control weights & 65.29 & 64.86 & 62.08 \\
%         New control weights & 66.94 & 66.94 & 66.94 \\ \hline \\
%         \textbf{Experimental treatment~2} \\ \hline
%         RW test statistic & 3.39\\
%         New test statistic & 3.48 \\ \hdashline
%         RW experimental treatment weights & 68.82 & 67.66 & 57.36 \\
%         New experimental treatment weights & 67.54 & 66.77 & 60.95 \\ \hdashline
%         RW control weights & 64.75 & 65.13 & 69.86 \\
%         New control weights & 62.96 & 62.96 & 62.96 \\ \hline
%     \end{tabular}
%     \caption{Example test statistics and weights for BAR scheme with $\mu_1 = \mu_2 = 0.5$. The standard `weights' (i.e.\ realised sample sizes per arm) would be 65, 67 and 63 for the control, treatment~1 and treatment~2, respectively.}
%     \label{tab:example1}
% \end{table}

\begin{table}[ht!]
    \centering
    \begin{tabular}{l l l l l}
        \textbf{Experimental treatment~1} \\ \hline
        RW test statistic & 1.11\\
        New test statistic & 1.13 \\ \hdashline
        RW experimental treatment weights &  61.38 & 60.11 & 71.09 \\
        New experimental treatment weights & 61.90 & 61.04 & 68.24 \\ \hdashline
        RW control weights & 65.59 & 66.05 & 62.90 \\
        New control weights & 65.44 & 65.44 & 65.44 \\ \hline \\
        \textbf{Experimental treatment~2} \\ \hline
        RW test statistic & 6.52\\
        New test statistic & 6.60 \\ \hdashline
        RW experimental treatment weights & 69.63 & 70.80 & 59.92 \\
        New experimental treatment weights & 68.07 & 68.84 & 62.84 \\ \hdashline
        RW control weights & 64.51 & 64.16 & 68.39 \\
        New control weights & 64.46 & 64.46 & 64.46 \\ \hline
    \end{tabular}
    \caption{Example test statistics and weights for BAR scheme with $\mu_1 = 0$ and $\mu_2 = 1$. The standard `weights' (i.e.\ realised sample sizes per arm) would be 65, 64 and 66 for the control, treatment~1 and treatment~2, respectively.}
    \label{tab:example1}
\end{table}

\begin{table}[ht!]
    \centering
    \begin{tabular}{l l l l l}
        \textbf{Experimental treatment~1} \\ \hline
        RW test statistic & -1.79\\
        New test statistic & -1.59 \\ \hdashline
        RW experimental treatment weights &  76.48 & 108.29 & 211.48 \\
        New experimental treatment weights & 69.00 & 86.05 & 130.34 \\ \hdashline
        RW control weights & 59.90 & 56.32 & 53.57 \\
        New control weights & 91.25 & 91.25 & 91.25 \\ \hline \\
        \textbf{Experimental treatment~2} \\ \hline
        RW test statistic & 2.92\\
        New test statistic & 1.66 \\ \hdashline
        RW experimental treatment weights & 53.67 & 35.71 & 8.16 \\
        New experimental treatment weights & 59.01 & 43.57 & 9.09 \\ \hdashline
        RW control weights & 72.21 & 98.47 & -62.83 \\
        New control weights & 14.26 & 14.26 & 14.26 \\ \hline
    \end{tabular}
    \caption{Example test statistics and weights for the error inflator scheme with $\mu_1 = 0, \mu_2 = 1$. The standard `weights' (i.e.\ realised sample sizes per arms) would be 65, 8 and 122 for the control, treatment~1 and treatment~2, respectively.}
    \label{tab:example2}
\end{table}

%\clearpage

\subsection{Simulation results}

To investigate the performance of the various approaches, we conducted  simulations for both the BAR and the error inflation scheme. As a standard comparison, we also provide simulation results for fixed (equal) randomization in the Supporting Information. The weighing approach from \cite{Robertson2019}, the proposal from Section \ref{sec_matching} and the naive approach (treating observed sample sizes as if they had been fixed in advance) were used. In all these approaches, the closed test procedure and the (Bonferroni-)Holm procedure are applied to adjust for the multiplicity arising from the testing of experimental treatments against a common control. In Tables \ref{tab:BAR_block} and \ref{tab:typeI_inflator}, \textit{disjunctive power} is the probability to reject at least one false null hypothesis (if there is one) and \textit{error} is the FWER. Nominal test levels are assumed to be $5\%$.

Table~\ref{tab:BAR_block} shows the results for the BAR scheme. As is well known, the closed test procedure has a slight power advantage if the treatments are equally effective, but is inferior when one of the treatments is effective, but the other(s) is not. FWER inflation did not occur in the simulations, even if the naive approach is used. The naive, the RW and the proposed approach lead to practically identical type~I errors and power here.

The results for the error inflator scheme are shown in Table \ref{tab:typeI_inflator}. We see that the error inflator scheme indeed does not control the FWER. The inflation remains somewhat modest, however, with a FWER not exceeding $7.5\%$ in any of the simulation scenarios. As expected, no FWER inflation arises when the two adaptive test methods are used. In line with what \cite{Robertson2019} observed, there is a price to pay for the FWER control: both methods tend to suffer from power losses relative to their naive counterparts. The proposed procedure, however, has higher power than the RW approach and for some scenarios, the gain is substantial. We speculate that this has to do with the fact that the variation of the weights is limited and cannot diverge as wildly from the observed sample sizes as they might with the RW approach (as illustrated in Table~\ref{tab:example2}).

% \begin{table}[ht!]

% \centering
% 	\caption{\label{tab:BAR_block_standard}  {Familywise error rate and disjunctive power for BAR using standard $z$-tests, for block randomization with a fixed control allocation. There were $10^5$ simulated trials for each set of parameter values.} \\}	

% \resizebox{0.7\linewidth}{!} {
% \begin{tabular}{p{0.5ex} l c c p{0cm} c c p{0cm} c c p{0cm} c c}
% 		& & \multicolumn{2}{c}{Closed $z$-test} & & \multicolumn{2}{c}{$z$-test (Holm)}
% 		& & \multicolumn{2}{c}{$z$-test (Bonferroni)} \\
% 		\cline{3-4} \cline{6-7} \cline{9-10} \\ [-2ex]
% 		& {Parameter values} & Error & Power & & Error & Power
% 		& & Error & Power \\[3pt] \hline
% 		\\
% 		1.\ & $\delta_1 = \delta_2 = 0$
% 		& 4.5 & - & & 4.3 & - & & 4.3 & -  \\ [1.5ex]
   		
%   		2.\ &  $\delta_1 = 0$, $\delta_2 = 0.5$
%   		& 4.8 & 61.7 & & 4.8 & 83.9 & & 2.4 & 83.9  \\ [1.5ex]
		
% 		3.\ & $\delta_1 = \delta_2 = 0.5$
%         & - &  {94.6} & & - & 92.3 & & - &  92.3   \\ [1.5ex]	
   		
% 		4.\ &  $\delta_1 = \delta_2 = \delta_3 = 0$
%   		& 3.7 & - & & 4.2 & - & & 4.2 & -   \\ [1.5ex]
   		   		   		
%   		5.\ &  $\delta_1 = \delta_2 = 0$, $\delta_3 = 0.5$
%   		& 4.5 & 35.7 & & 4.4 &  71.4 & & 3.0 & 71.3  \\ [1.5ex]
   		
%   		6.\ &  $\delta_1 = 0$, $\delta_2 = \delta_3 = 0.5$
% 		& 4.9 & 66.7 & & 4.5 &  85.3 & & 1.7 & 85.2 \\ [1.5ex]
		
% 		7.\ &  $\delta_1 = 0$, $\delta_2 = 0.25$, $\delta_3 = 0.5$
% 		& 4.5 & 50.6 & & 3.6 &  72.3 & & 1.6 & 72.3  \\ [1.5ex]

%   		8.\ &  $\delta_1 = \delta_2 = \delta_3 = 0.5$
%   		& - &  93.2 & &  - & 90.3 & & - & 90.3 \\[6pt] \hline

% 		\end{tabular}
% 		}
% \end{table}

\begin{sidewaystable}[ht!]

\resizebox{\linewidth}{!} {
\begin{tabular}{p{0.5ex} l c c p{0cm} c c p{0cm} c c p{0cm} c c p{0cm} c c p{0cm} c c}
		& & \multicolumn{2}{c}{Closed $z$-test} & & \multicolumn{2}{c}{RW closed test}
		& & \multicolumn{2}{c}{New closed test} & & \multicolumn{2}{c}{$z$-test (Holm)}
		& & \multicolumn{2}{c}{RW test (Holm)}  & &  \multicolumn{2}{c}{New test (Holm)} \\
		\cline{3-4} \cline{6-7} \cline{9-10} \cline{12-13}
		\cline{15-16} \cline{18-19}\\ [-2ex]
		& {Parameter values} & Error & Power & & Error & Power & & Error & Power & & Error & Power & & Error & Power & & Error & Power \\[3pt] \hline
		\\
		1.\ & $\delta_1 = \delta_2 = 0$
		& 4.5 & - & & 4.5 & - & & 4.5 & - & & 4.3 & - & & 4.4 & - & & 4.4 & -    \\ [1.5ex]
   		
   		2.\ &  $\delta_1 = 0$, $\delta_2 = 0.5$
   		& 4.8 & 61.7 & & 4.9 & 61.7 & & 5.0 & 61.7 & & 4.8 & 83.9 & & 4.9 & 83.8 & & 4.9 & 83.9 \\ [1.5ex]
		
		3.\ & $\delta_1 = \delta_2 = 0.5$
        & - &  {94.6} & & - & 94.6 & & - & 94.6 & & - & 92.3 & & - & 92.4 & & - & 92.4 \\ [1.5ex]	
   		
		4.\ &  $\delta_1 = \delta_2 = \delta_3 = 0$
   		& 3.7 & - & & 3.8 & - & & 3.8 & - & & 4.2 & - & & 4.5 & - & & 4.5 & - \\ [1.5ex]
   		   		   		
   		5.\ &  $\delta_1 = \delta_2 = 0$, $\delta_3 = 0.5$
   		& 4.5 & 35.7 & & 4.5 & 35.7 & & 4.5 & 35.7 & & 4.4 & 71.4 & & 4.6 & 71.4 & & 4.5 & 71.3 \\ [1.5ex]
   		
   		6.\ &  $\delta_1 = 0$, $\delta_2 = \delta_3 = 0.5$
		& 4.9 & 66.7 & & 5.0 & 67.1 & & 5.0 & 67.0 & & 4.5 & 85.3 & & 4.7 & 85.4 & & 4.7 & 85.3 \\ [1.5ex]
		
		7.\ &  $\delta_1 = 0$, $\delta_2 = 0.25$, $\delta_3 = 0.5$
		& 4.5 & 50.6 & & 4.6 & 50.8 & & 4.6 & 50.7 & & 3.6 & 72.3 & & 3.7 & 72.5 & & 3.7 & 72.4 \\ [1.5ex]

   		8.\ &  $\delta_1 = \delta_2 = \delta_3 = 0.5$
   		& - &  93.2 & &  - & 93.3 & & - & 93.3 & & - & 90.3 & & - & 90.6 & & - & 90.5 \\[6pt] \hline

		\end{tabular}
		}
		
		\vspace{12pt}
		
		\caption{\label{tab:BAR_block}  Familywise error rate and disjunctive power for BAR using block randomization with a fixed control allocation. There were $10^5$ simulated trials for each set of parameter values.}

\end{sidewaystable}

\begin{sidewaystable}[ht!]

\resizebox{\linewidth}{!} {
\begin{tabular}{p{0.5ex} l c c p{0cm} c c p{0cm} c c p{0cm} c c p{0cm} c c p{0cm} c c}
		& & \multicolumn{2}{c}{Closed $z$-test} & & \multicolumn{2}{c}{RW closed test}
		& & \multicolumn{2}{c}{New closed test} & & \multicolumn{2}{c}{$z$-test (Holm)}
		& & \multicolumn{2}{c}{RW test (Holm)}  & &  \multicolumn{2}{c}{New test (Holm)} \\
		\cline{3-4} \cline{6-7} \cline{9-10} \cline{12-13}
		\cline{15-16} \cline{18-19}\\ [-2ex]
		& {Parameter values} & Error & Power & & Error & Power & & Error & Power & & Error & Power & & Error & Power & & Error & Power \\[3pt] \hline
		\\
		1.\ & $\delta_1 = \delta_2 = 0$
		& 4.8 & - & & 4.0 & - & & 4.3 & - & & \textbf{6.3} & - & & 4.8 & - & & 5.0 & -    \\ [1.5ex]
   		
   		2.\ &  $\delta_1 = 0$, $\delta_2 = 1$
   		& \textbf{7.3} & 26.8 & & 5.0 & 24.0 & & 4.9 & 25.5 & & \textbf{7.2} & 81.0 & & 4.1 & 44.8 & & 4.3 & 57.6 \\ [1.5ex]
		
		3.\ & $\delta_1 = \delta_2 = 0.5$
        & - &  94.7 & & - & 93.0 & & - & 93.8 & & - & 92.0 & & - & 88.4 & & - & 90.0 \\ [1.5ex]	
   		
		4.\ &  $\delta_1 = \delta_2 = \delta_3 = 0$
   		& 4.0 & - & & 3.2 & - & & 3.5 & - & & \textbf{5.8} & - & & 4.7 & - & & 4.9 & - \\ [1.5ex]
   		   		   		
   		5.\ &  $\delta_1 = \delta_2 = 0$, $\delta_3 = 1$
   		& 4.7 & 19.4 & & 3.8 & 16.4 & & 4.0 & 18.1 & & \textbf{6.1} & 76.1 & & 4.3 & 46.2 & & 4.5 & 56.4 \\ [1.5ex]
   		
   		6.\ &  $\delta_1 = 0$, $\delta_2 = \delta_3 = 1$
		& \textbf{7.3} & 25.2 & & 4.9 & 25.9 & & 4.9 & 26.5 & & \textbf{7.0} & 91.9 & & 3.8 & 61.4 & & 4.1 & 75.2 \\ [1.5ex]
		
		7.\ &  $\delta_1 = 0$, $\delta_2 = 0.5$, $\delta_3 = 1$
		& \textbf{7.2} & 23.0 & & 4.7 & 22.0 & & 4.8 & 23.2 & & \textbf{6.1} & 80.1 & & 3.3 & 50.0 & & 3.6 & 61.4 \\ [1.5ex]

   		8.\ &  $\delta_1 = \delta_2 = \delta_3 = 0.5$
   		& - & 93.8 & &  - & 91.6 & & - & 92.7 & & - & 89.8 & & - & 84.5 & & - & 86.6 \\[6pt] \hline

		\end{tabular}
		}
		
    \vspace{12pt}

	\caption{\label{tab:typeI_inflator} Familywise error rate and disjunctive power for the error inflator scheme using block randomization with a fixed control allocation. There were $10^5$ simulated trials for each set of parameter values.}			

\end{sidewaystable}

\section{Discussion}
\label{sec:discuss}

%Optimal design and potential trade-off between different objectives. RAR vs Fixed/Equal Randomisation. Binary endpoints. Control adaptation.

In this paper, we have proposed an improved testing strategy based on the one by~\cite{Robertson2019}, which guarantees FWER control in the context of block-randomised response-adaptive trials with a fixed control allocation. Our proposal is simpler but is more restrictive as it is not applicable to fully sequential RAR or having an adaptive control allocation. However, our proposal guarantees that the weights are non-negative, and there can be substantial power gains in some settings.

As noted in~\cite{Robertson2019}, since the proposed testing procedure is based on the CIV principle, it has the additional important flexibility of being valid when the allocation is changed due to external information. Our proposal is also designed for normally-distributed outcomes, although it can apply for other types of outcomes through the use of asymptotics. However, a natural extension of this work would be to work directly with binary endpoints (for example) and potentially apply the CIV principle to this setting.

The use of the CIV principle to ``reweight'' the test statistics raises interesting questions around the design of optimal response-adaptive trials (i.e.\ the formulation of RAR procedures that optimise certain criteria). For example, some RAR procedures incorporate a formal power constraint, but this is based on standard test statistics. If an alternative testing strategy such as our proposed one is used, then there is a mismatch between the optimality criterion and the subsequent analysis of the trial.

More generally, it is important to remember that there can be trade-offs between the different objectives in a trial. Indeed, we have seen that guaranteeing FWER control power can lead to a substantial loss in power. As another example, more `extreme' RAR procedures (i.e.\ those that skew the randomization probabilities close to 0 or 1) that perform well in terms of patient benefit metrics may conversely have low power. Hence the question of whether to use RAR as opposed to a fixed randomisation scheme is not a simple one, and crucially depends on the trial context and goals.

\section*{Acknowledgements}

DSR was funded by the UK Medical Research Council (MC\_UU\_00002/6) and the Biometrika Trust. The views expressed in this publication are those of the authors and not necessarily those of the NHS, the National Institute for Health Research or the Department of Health and Social Care (DHSC). %\\
\noindent This work was funded by UKRI grant MC\_UU\_00002/6. For the purpose of open access, the author has applied a Creative Commons Attribution (CC BY) licence to any Author Accepted Manuscript version arising. \\[-6pt]

\noindent \textbf{Data availability}: Data sharing is not applicable to this article as no datasets were generated or analysed.

\bibliography{references}

\clearpage

\appendix

\section*{Supporting information}

\section{Details of RAR procedures}

\subsection{Bayesian Adaptive Randomization (BAR)}

Recall that the efficacy outcome for the $i$th treatment follows a $N(\mu_i, 1)$ distribution. We assign independent normal priors to the~$\mu_i$ ($i = 0, 1, \ldots, K$), such that $\mu_i \sim N(\mu_{i,0} , \sigma_{i,0}^2)$. Let $D_j$ denote the total number of patients allocated to the experimental treatments by the end of the $j$th block, and $\tilde{n}_{i,j}$ denote the number of patients allocated to treatment~$i$ by the end of the $j$th block.
At stage~$(j+1)$, when the outcomes $\bm{x} = (x_1, \ldots,  x_{D_j})$ have been observed, the posterior for~$\mu_i$ is as follows: \[
\mu_i \mid \bm{X} = \bm{x} \sim N\left( \frac{\sigma_{i,0}^2}{1 + \tilde{n}_{i,j}\sigma_{i,0}^2} \sum_{k=1}^{D_j}\mathds{1}_{\{a_k = i\}} x_k + \frac{\tilde{n}_{i,j}}{1+\tilde{n}_{i,j} \sigma_{i,0}^2}\mu_{i,0} \;  , \frac{\sigma_{i,0}^2}{1+\tilde{n}_{i,j} \sigma_{i,0}^2} \right).
\]
\noindent In our simulations, for simplicity we set the priors $\mu_{i,0} = 0$ and $\sigma_{i,0}^2 = 1$, while $\gamma = 0.5$.

\subsection{Error inflator scheme}

Using the same notation as above, the allocation probabilities for block $j  \in \{ 1, \ldots, J-1 \}$, patient $k =  D_j + 1, \ldots, D_{j+1}$ and treatment $l \in \{2, \ldots, K\}$ are: \begin{align*}
P(a_k = 1) & = \begin{cases}
0 & \text{if } \; \sum_{i = 1}^{D_j} \mathds{1}_{\{a_i = 1\}} \frac{X_i}{\tilde{n}_{1,j}} > 0.5 \\
1 & \text{otherwise}
\end{cases} \\
P(a_k = l) & = \begin{cases}
1/K & \text{if } \; \sum_{i = 1}^{D_j} \mathds{1}_{\{a_i = 1\}} \frac{X_i}{\tilde{n}_{1,j}} > 0.5 \\
0 & \text{otherwise}
\end{cases}
\end{align*}

\section{Fixed randomisation simulation study}

Table~\ref{tab:fixed} shows the familywise error rate and disjunctive power for fixed (equal) randomization. %There were $10^5$ simulated trials for each set of parameter values.

\setcounter{table}{0}
\renewcommand{\thetable}{B\arabic{table}}

\begin{sidewaystable}[ht!]

\resizebox{\linewidth}{!} {
\begin{tabular}{p{0.5ex} l c c p{0cm} c c p{0cm} c c p{0cm} c c p{0cm} c c p{0cm} c c}
		& & \multicolumn{2}{c}{Closed $z$-test} & & \multicolumn{2}{c}{RW closed test}
		& & \multicolumn{2}{c}{New closed test} & & \multicolumn{2}{c}{$z$-test (Holm)}
		& & \multicolumn{2}{c}{RW test (Holm)}  & &  \multicolumn{2}{c}{New test (Holm)} \\
		\cline{3-4} \cline{6-7} \cline{9-10} \cline{12-13}
		\cline{15-16} \cline{18-19}\\ [-2ex]
		& {Parameter values} & Error & Power & & Error & Power & & Error & Power & & Error & Power & & Error & Power & & Error & Power \\[3pt] \hline
		\\
		1.\ & $\delta_1 = \delta_2 = 0$
		& 4.7 & - & & 4.6 & - & & 4.6 & - & & 4.6 & - & & 4.6 & - & & 4.5 & -    \\ [1.5ex]
   		
   		2.\ &  $\delta_1 = 0$, $\delta_2 = 0.5$
   		& 5.0 & 50.0 & & 5.0 & 50.0 & & 5.0 & 50.0 & & 5.0 & 81.5 & & 5.0 & 81.4 & & 4.9 & 81.2 \\ [1.5ex]
		
		3.\ & $\delta_1 = \delta_2 = 0.5$
        & - &  {94.6} & & - & 94.6 & & - & 94.5 & & - & 92.3 & & - & 92.2 & & - & 92.1 \\ [1.5ex]	
   		
		4.\ &  $\delta_1 = \delta_2 = \delta_3 = 0$
   		& 3.7 & - & & 3.7 & - & & 3.7 & - & & 4.6 & - & & 4.6 & - & & 4.6 & - \\ [1.5ex]
   		   		   		
   		5.\ &  $\delta_1 = \delta_2 = 0$, $\delta_3 = 0.5$
   		& 4.5 & 27.6 & & 4.5 & 27.5 & & 4.5 & 27.5 & & 4.6 & 67.1 & & 4.6 & 66.9 & & 4.6 & 66.7 \\ [1.5ex]
   		
   		6.\ &  $\delta_1 = 0$, $\delta_2 = \delta_3 = 0.5$
		& 5.0 & 55.7 & & 5.0 & 55.7 & & 5.0 & 55.7 & & 4.7 & 83.6 & & 4.7 & 83.5 & & 4.7 & 83.4 \\ [1.5ex]
		
		7.\ &  $\delta_1 = 0$, $\delta_2 = 0.25$, $\delta_3 = 0.5$
		& 4.6 & 42.6 & & 4.6 & 42.5 & & 4.6 & 42.5 & & 3.8 & 69.8 & & 3.8 & 69.7 & & 3.8 & 69.5 \\ [1.5ex]

   		8.\ &  $\delta_1 = \delta_2 = \delta_3 = 0.5$
   		& - &  93.2 & &  - & 93.1 & & - & 93.1 & & - & 90.1 & & - & 90.0 & & - & 90.0 \\[6pt] \hline

		\end{tabular}
		}
		
		\vspace{12pt}
		
		\caption{\label{tab:fixed} Familywise error rate and disjunctive power for fixed (equal) randomization. There were $10^5$ simulated trials for each set of parameter values.}	

\end{sidewaystable}

\end{document}